\documentclass[prl,aps,twocolumn,preprintnumbers]{revtex4}
\usepackage{epsfig,amssymb,amsfonts}

\begin{document}
\title{Photon blockade induced Mott transitions
and XY spin models in coupled cavity arrays}
    \author{Dimitris G. \surname{Angelakis}$^{1}$}
    \email{dimitris.angelakis@qubit.org}
\author{Marcelo F. \surname{Santos}$^{2}$}%
\author{Sougato Bose $^{3}$}
\address{$^{1}$Centre for Quantum Computation, Department
of Applied Mathematics
 and Theoretical Physics, University of Cambridge,
 Wilberforce Road, CB3 0WA, UK}
\address{$^{2}$Dept. de F\'{\i}sica, Universidade Federal de
  Minas Gerais, Belo Horizonte,
30161-970, MG, Brazil}
  \address{$^{3}$Department of Physics and Astronomy, University
College London, Gower St., London WC1E 6BT, UK}


\begin{abstract} As photons do not interact with each other, it is interesting to
ask whether photonic systems can be  modified to exhibit the phases
characteristic of strongly coupled many-body systems. We demonstrate
how a Mott insulator type of phase of excitations can arise in an
array of coupled electromagnetic cavities, each of which is coupled
resonantly to a {\em single} two level system (atom/quantum
dot/Cooper pair) and can be individually addressed from outside. In
the Mott phase each atom-cavity system has the same integral number
of net polaritonic (atomic plus photonic) excitations with photon
blockade providing the required repulsion between the excitations in
each site. Detuning the atomic and photonic frequencies suppresses
this effect and induces a transition to a photonic superfluid. We
also show that for zero detuning, the system can simulate the
dynamics of many body spin systems.
\end{abstract}

\maketitle
\bibliographystyle{apsrev}
 \preprint{quant-ph/0606159-20/6/06}

{\em Introduction:} Strongly coupled many-body systems described by
the Bose-Hubbard models \cite{fisher} exhibit Mott insulating phases
whose realization in optical lattices \cite{jaksch,greiner,kasevich}
have opened varied possibilities for the simulation of many body
physics \cite{Duan-Lukin-Demler-PRL}. Can there be another {\em
engineered} quantum many-body system which displays such phases?
This will be especially interesting if the strengths of this system
are ``complementary" to that of optical lattices -- for example if
it allowed the {\em co-existence} of accessibility to individual
constituents of a many-body system and a strong interaction between
them, or if it allowed the simulation of arbitrary networks rather
than those derivable from superposing lattices. Particularly
arresting will be to find such phases by {\em minimally modifying} a
system of photons which, by being non-interacting, are unlikely
candidates for the studies of many-body phenomena. Here we propose
such a system consisting of coupled electromagnetic
cavities\cite{haroche,grangier,yariv,orzbay,krauss,trupke}, doped
with {\em single} two level
systems\cite{coupsupercond,vuckovic,toroid,mabuchi,badolato,
supercond1,supercond2,song,kimble-latest}. Using the nonlinearity
generated from the corresponding photon blockade
effect\cite{blockade,blockade1}, we show the possibility of
observing an insulator phase of total (atomic$+$photonic or simply
{\em polaritonic}) excitations and its transition to a superfluid of
photons. Compared to the optical lattices case, the different
Hubbard-like model which describes the system involves {\em neither
purely bosonic nor purely fermionic} entities, the transition from
insulator to superfluid is also accompanied by a transition of the
excitations from polaritonic to photonic, and that {\em
excitations}, rather than physical particles such as atoms are
involved. In addition, the possibility to simulating the dynamics of
an XY spin chain-with individual spin manipulation, using a
mechanism different from that used in optical lattices
\cite{Duan-Lukin-Demler-PRL} is suggested.

{\em System Description:} Assume a chain of $N$ coupled cavities. A
realization of this has been studied in structures known as a
coupled resonator optical waveguide (CROWs) or couple cavities
waveguides(CCW) in photonic
crystals\cite{orzbay,krauss,angelakis,yariv}, in tabered fiber
coupled toroidal microcavies\cite{toroid} and coupled
superconducting microwave resonators
\cite{coupsupercond,supercond1,supercond2}. This has already
stimulated proposals to use such systems to implement optical and
cluster state quantum computing\cite{angelakis04,angelakis-kay07a},
produce entangled photons\cite{angelakis06} and study interacting
polaritons\cite{plenio}. We will describe the system dynamics using
the operators corresponding to the localized eigenmodes (Wannier
functions), $a^{\dagger}_{k}(a_{k})$. The Hamiltonian is given by$
H=\sum_{k=1}^{N}\omega_d
a^{\dagger}_{k}a_{k}+\sum_{k=1}^{N}A(a^{\dagger}_{k}a_{k+1}+H.C.)$
 and corresponds to a series quantum harmonic oscillators
coupled through hopping photons. The photon frequency and hopping
rate is $\omega_{d}$ and $A$ respectively and no nonlinearity is
present yet. Assume now that the cavities are doped with two level
systems (atoms/ quantum dots/superconducting qubits) and
$|g\rangle_{k}$ and $|e\rangle_{k}$ their ground and excited states
at site $k$. The Hamiltonian describing the system is the sum of
three terms. $H^{free}$ the Hamiltonian for the free light and
dopant parts, $H^{int}$ the Hamiltonian describing the internal
coupling of the photon and dopant in a specific cavity and $H^{hop}$
for for the light hopping between cavities.
\begin{eqnarray}
H^{free}&&=\omega_{d}\sum_{k=1}^N a_k^\dagger a_k+\omega_{0}\sum_k|e\rangle_{k} \langle e|_{k} \\
H^{int}&&=g \sum_{k=1}^N(a_k^\dagger|g\rangle_{k}\langle e|_{k}+H.C.)\\
H^{hop}&&= A\sum_{k=1}^N(a_k^\dagger a_{k+1} +H.C)
\end{eqnarray}
 where g is the light atom coupling strength. The $H^{free}+H^{int}$ part of the
Hamiltonian can be diagonalized
 in a basis of mixed photonic and atomic excitations, called
{\it polaritons}. These polaritons, also known as dressed states,
involve an equal mixture of photonic and atomic excitations and are
defined on resonance by creation operators
$P_{k}^{(\pm,n)\dagger}=(|g,n\rangle_k\langle g,0|_k\pm
|e,n-1\rangle_k\langle g,0|_k)/\sqrt2$, where $|n\rangle_k,
|n-1\rangle_k$ and $|0\rangle_k$ denote $n, n-1$ and $0$ photon Fock
states in the $k$th cavity. The polaritons of the $k$th atom-cavity
system are denoted as $|n\pm\rangle_k$ and given by
$|n\pm\rangle_k=(|g,n\rangle_k\pm |e,n-1\rangle_k)/\sqrt2$ with
energies $E^{\pm}_{n}=n\omega_{d}\pm g\sqrt{n}$ and are also
eigenstates of the the sum of the photonic and atomic excitations
operator ${\cal N}_k=a_k^\dagger a_k+|e\rangle\langle e|_k$ with
eigenvalue $n$(Fig. 1).

\begin{figure}

    \includegraphics[width=7cm]{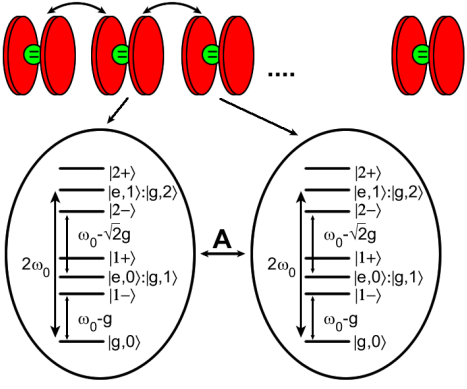}

\caption{A series of coupled cavities coupled through light and the
polaritonic energy levels for two neighbouring cavities. }
\label{pol}
\end{figure}

{\it Polaritonic Mott State: } We will now justify that 
the lowest energy
states of the system consistent with a given number of net
excitations per site (or filling factor) becomes a Mott state of the
net (polaritonic) excitations for integer values of the filling
factor. To understand this, we rewrite the Hamiltonian (for
$\Delta\sim 0$) in terms of the polaritonic operators (defined in
the caption of Fig. 1) as
\begin{eqnarray}
H=\sum_{k=1}^{N}[\sum_{n=1}^{\infty}n(\omega_{d}-g)P_{k}^{(-,n)\dagger}P_{k}^{(-,n)}+\nonumber\\
\sum_{n=1}^{\infty}n(\omega_{d}+g)P_{k}^{(+,n)\dagger}P_{k}^{(+,n)}+\nonumber\\
\sum_{n=1}^{\infty}g(n-\sqrt{n})P_{k}^{(-,n)\dagger}P_{k}^{(-,n)}+\nonumber\\
\sum_{n=1}^{\infty}g(\sqrt{n}-n)P_{k}^{(+,n)\dagger}P_{k}^{(+,n)}]+\nonumber\\
A\sum_{k=1}^{N}(a_k^\dagger a_{k+1} +H.C). \label{nodet}
\end{eqnarray}
The above implies (assuming the regime $A<<g\sqrt{n}<<\omega_d$)
that the lowest energy state for a given number, say $\eta$, of net
excitations at the $k$th site would be the state $|\eta-\rangle_k$
(this is because $|\eta+\rangle$ has a higher energy, but same net
excitation $\eta$). Thus one need only consider the first, third and
last lines of the above Hamiltonian $H$ for determining the lowest
energy states. The first line corresponds to a linear spectrum,
equivalent to that of a harmonic oscillator of frequency
$\omega_{d}-g$. If only that part was present in the Hamiltonian,
then it would not cost any extra energy to add an excitation (of
frequency $\omega_{d}-g$) to a site already filled with one or more
excitations, as opposed to an empty site. However, the term
$g(n-\sqrt{n})P_{k}^{(-,n)\dagger}P_{k}^{(-,n)}$ raises energies of
uneven excitation distribution such as
$|(n+1)-\rangle_k|(n-1)-\rangle_{l}$
 among any two sites $k$ and $l$ relative to the uniform excitation distribution
$|n-\rangle_k|n-\rangle_{l}$ among these sites. Thus the third line
of the above Hamiltonian can be regarded as an effective, nonlinear
``on-site" {\it photonic repulsion}, and leads to a Mott state of
the net excitations per site being the ground state for commensurate
filling.  Reducing the strength of the effective nonlinearity, i.e.,
the blockade effect, should drive the system to the superfluid
regime. For this, one should move the system away from the strong
resonant interaction to a weaker dispersive regime. This could be
done by Stark shifting and detuning (globally again) the atomic
transitions from the cavity by an external field. The new detuned
polaritons are not as well separated as before and their energies
are merely shifts of the bare atomic and photonic ones by $\pm
g^2(n+1)/\delta$ for the $|e,n\rangle$ and $|g,n+1\rangle$
respectively. In this case it costs no extra energy to add
excitations (excite transitions to higher polaritons) in a single
site, and the system moves to the superfluid regime.

\begin{figure}

    \includegraphics[width=9cm]{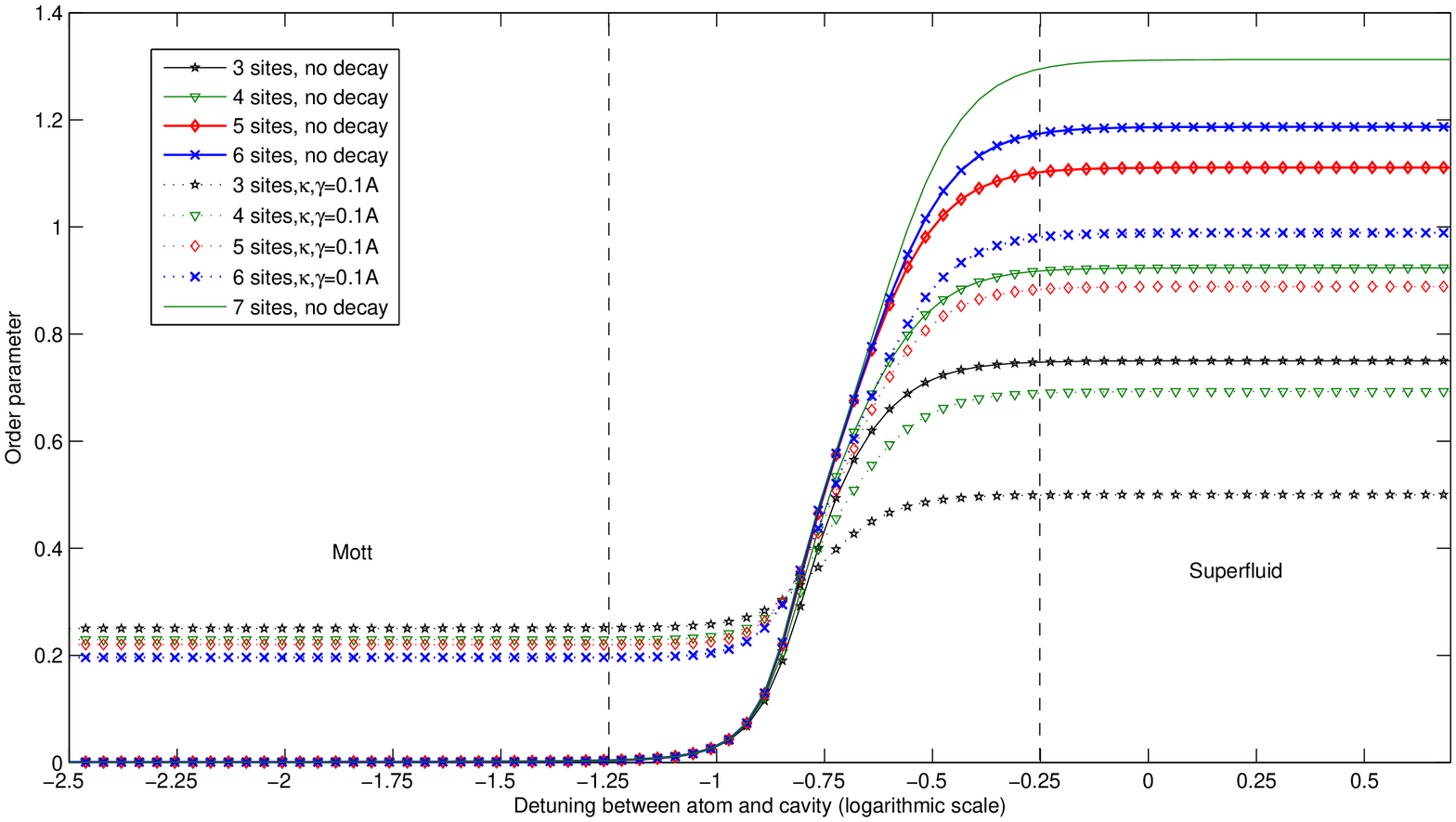}

\caption{The order parameter as a function of the detuning  between
the hopping photon and the doped two level system(in logarithimic
units of the matter-light coupling g). Simulations include results
for 3-7 sites, with and without dissipation due to spontaneous
emission and cavity leakage. Close to resonance ($0 \le \Delta \le
10^{-1})$, where the photon blockade induced nonlinearity is maximum
(and much larger than the hopping rate), the system is forced into a
polaritonic Fock state with same integral number of excitations per
site(order parameter zero-Mott insulator state). Detuning the system
by applying external fields and inducing Stark shifts ($\Delta \ge
g$), weakens the blockade and leads to the appearance of different
coherent superpositions of excitations per site( a photonic
superfluid).The increase in number of sites leads to a sharper
transition as expected. } \label{var}
\end{figure}

To justify the transition of the system from a Mott phase to a
superfluid phase as detuning $\Delta=\omega_0-\omega_d$ is
increased, we have performed a numerical simulation using 3-7 sites
using the Hamiltonian of Eq.(1)-(3)(numerical diagonalization of the
complete Hamiltonian without any approximations ) \footnote{Finite
numbers of sites is often used in studying the physics of the
transition from Mott to superfluid phases \cite{jaksch,burnett},
especially as analytic methods from mean-field theory are invalid
for one-dimensional or network geometries. In the pioneering optical
lattice Mott transition paper \cite{jaksch}, for example, 7 lattice
sites were used, and we use the same though our model is
computationally more exhaustive than just having bosonic occupation
numbers per site because of the extra atomic degree of freedom at
each site that  cannot be eliminated.}. In the Mott phase the
particle number per site is fixed and its variance is zero (every
site is in a Fock state). In such a phase, the expectation value of
the destruction operator for the relevant particles-the order
parameter is zero. In the traditional mean field (and thus
necessarily approximate) picture, this expectation value becomes
finite on transition to a superfluid, as a {\em coherent}
superposition of different particle numbers is allowed to exist per
site. However, our entire system is a ``closed" system and there is
no particle exchange with outside. Superfluid states are
characterized by a fixed ``total" number of particles in the three
site system and the expectation of a destruction operator in any
given site is zero even in the superfluid phase. Thus this
expectation value cannot be used as an order parameter for a quantum
phase transition. Instead we use the variance of total number of
excitations per site, the operator ${\cal N}_k$, in a given site (we
choose the middle cavity, but any of the other cavities would do) to
characterize the Mott to superfluid phase transition. This variance
$var({\cal N}_k)$ has been plotted in Fig.\ref{var} as a function of
$\log_{10}\Delta$ for a filling factor of one net excitation per
site. For this plot, we have taken the parameter ratio $g/A=10^{2}$
($g/A=10^{1}$ gives very similar results), with $\Delta$ varying
from $\sim 10^{-3}g$ to $\sim g$ and $\omega_d,\omega_0 \sim 10^4g$.
We have plotted both ideal graphs (if neither the atoms nor the
cavity fields underwent any decay or decoherence) and also performed
simulations {\em explicitly} using decay of the atomic states and
photonic states in the range of $g/max(\kappa,\gamma)\sim 10^3$,
where $\kappa$ and $\gamma$ are cavity and atomic decay rates.

 These decay rates are
expected soon to be feasible in toroidal microcavity systems with
atoms \cite{toroid} and arrays of coupled stripline microwave
resonators, each interacting with a superconducting qubit
\cite{supercond2}. For these simulations we have assumed that the
experiment (of going from the Mott state to the superfluid state and
back) takes place in a time-scale of $1/A$ so that the evolution of
one ground state to the other and back is adiabatic. The simulations
of the state with decay have been done using quantum jumps, and it
is seen that there is {\em still a large difference of $var({\cal
N}_k)$ between the Mott and superfluid phases despite the decays}.
As expected the effect of dissipation reduces the final value of
order parameter in the superfluid regime(population has been lost
through decay) whereas in the Mott regime leads to the introduction
of fluctations again due to population loss from the $|1-\rangle$
state. The Mott ($var({\cal N}_k)=0$) to superfluid ($var({\cal
N}_k)=0.75$) transition takes place over a finite variation of
$\Delta$ (because of the finiteness of our lattice) around $10^2g$
and as expected becomes sharper as the number of sites is increased.

\begin{figure}
\includegraphics[height=4cm]{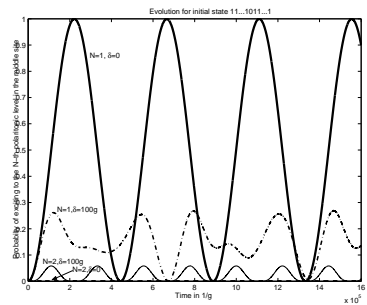}

\caption{Up: The probability of exciting the polaritons
corresponding to one (thick solid line) or two (thick dashed line)
excitations for the middle cavity which taken initially to be empty(
in the resonant regime). In this case, as we started with fewer
polaritons than number of cavities, oscillations occur for the
single excitation polariton. The double one is never occupied due to
the blockade effect-Mott insulator phase(thick dashed line). For the
detuned case however, hopping of more than one polaritons is
allowed(superfluid regime) which is evident as the higher
polaritonic manifolds are being populated now(dashed line and thin
solid line).} \label{pop}
\end{figure}

   In an experiment one would start in the resonant (Mott) regime with
all atom-cavity systems initially in their absolute ground
$(|g,0\rangle^{\otimes k})$ states and prepare the atom-cavity
systems in the joint state $|1-\rangle^{\otimes k}$ by applying a
global external laser tuned to this transition. This is the Mott
state with the total (atomic+photonic) excitations operator ${\cal
N}_k$ having the value unity at each site. One would then Stark
shift and detune (globally again) the atomic transitions from the
cavity by an external field and observe the probability finding
$|1-\rangle$  and the predicted decrease of this probability
(equivalent to the increase in our order parameter, the variance of
${\cal N}_k$) as the detuning is increased. For inferring the
fluctuations in ${\cal N}_k$ of our system, it is basically suffices
to check the population of the $|1-\rangle$ state, as this is an
eigenstate of ${\cal N}_k$. For this, a laser is applied which is of
the right frequency to accomplish a cycling transition between
$|1-\rangle$ and another probe third level. Through monitoring its
fluorescence, accurate state measurements can be made in the
standard atomic state measurement way\cite{Rowe00}. 

To strengthen our case, we have also calculated the probabilities of
populating the lowest polaritonic state of the first and second
manifolds $|1-\rangle_k$, $|2-\rangle_k$ of a middle cavity out of
an array of 3,5 and 7 cavities. The initial state is the polaritonic
state $|1-\rangle_k$ excited at the right and the left cavities,
with the middle cavity being in the ground state $|g,0\rangle_k$.
Figure \ref{pop} shows as expected from our discussion above, on
resonance the photon blockade is preventing any excitation to any
state higher than the first({\it Mott insulator phase}). However, by
simply varying the atomic frequency and inducing some detuning (of
the order of $100g$ in our simulation), the weakening of the
blockade effect results in the probability of exciting to the second
manifold to become increasingly strong. This means that the
polaritonic excitations(or better the photon like polaritonic
particles now as we are in the dispersive regime) can hop together
in bunches of two or more from cavity to cavity({\it superfluid
regime}).


{\it Simulating spin models: }We will now show that in the Mott
regime the system simulates a XY spin model with the presence and
absence of polaritons corresponding to spin up and down. Let us
assume we initially populate the lattice only with polaritons of
energy $\omega_{0}-g$, in the limit $\omega_{d}\approx\omega_{0}$,
Eqs. (2)-(4) becomes
\begin{eqnarray}
H_{k}^{free}=\omega_{d}\sum_{k=1}^{N}&&P^{(+)\dagger}_{k}P^{(+)}_{k}+
P^{(-)\dagger}_{k}P^{(-)}_{k}\\
H_{k}^{int}=g\sum_{k=1}^{N}&&P^{(+)\dagger}_{k}P^{(+)}_{k}-P^{(-)\dagger}_{k}P^{(-)}_{k}\\
H_{k}^{hop.}=A\sum_{k=1}^{N}&&P^{(+)\dagger}_{k}P^{(+)}_{k+1}+P^{(-)\dagger}_{k}P^{(+)}_{k+1}
+\nonumber\\
&&P^{(+)\dagger}_{k}P^{(-)}_{k+1}+P^{(-)\dagger}_{k}P^{(-)}_{k+1}+H.C.
\label{H_hop_pol}
\end{eqnarray}
where $P_{k}^{(\pm)\dagger}=P_{k}^{(\pm,1)\dagger}$ is the
polaritonic operator creating excitations to the first polaritonic
manifold (Fig. 1). In the rotating wave approximation, Eq.
\ref{H_hop_pol} reads (in the interaction picture).
$H_{I}=A\sum_{k=1}^{N}P^{(-)\dagger}_{k}P^{(-)}_{k+1}+H.C.$
In deriving the above, the logic is two step. Firstly note that the
terms of the type $P^{(-)\dagger}_{k}P^{(+)}_{k+1}$, which
inter-convert between polaritons, are fast rotating and they vanish.
Secondly, if we create only the polaritons $P^{(-)\dagger}_{k}$ in
the lattice initially with energy $\omega_{0}-g$, then the
polaritons corresponding to $P^{(+)\dagger}_{k}$ will never even be
created, as the inter-converting terms are vanishing. Thus the term
$P^{(+)\dagger}_{k}P^{(+)}_{k}$ can also be omitted. Note that
because the double occupancy of the sites is prohibited, one can
identify $P^{(-)\dagger}_{k}$ with
$\sigma^{+}_k=\sigma^x_k+i\sigma^y_k$, where $\sigma^x_k$ and
$\sigma^y_k$ are standard Pauli operators. Then the Hamiltonian
becomes $ H_I=\sigma^x_k\sigma^x_{k+1}+\sigma^y_k\sigma^y_{k+1}$.
The latter is the standard XY model of interacting spins with spin
up/down corresponding to the presence/absence of a polariton. Note
that although this is different with optical lattice realizations of
spin models, where instead, the internal levels of a two level atom
are used for the two qubit states \cite{Duan-Lukin-Demler-PRL}, the
measurement could be done using very similar atomic state
measurement techniques (utilizing the advantage of larger distances
between sites as well). Some simple applications of XY spin chains
in quantum information processing such as quantum information
transfer \cite{bose,giovannetti,bose-burgarth07a} can thus be
readily implemented in our system. Very recently a novel idea for
efficient cluster state quantum computation was proposed in this
system where the database search and factoring quantum algorithms
could be implemented using just two rows of
cavities\cite{angelakis-kay07a}.

In conclusion we showed that a range of many body system effects,
such Mott transitions for polaritonic particles obeying mixed
statistics could be observed in optical systems of individual
addresable coupled cavity arrays interacting with two level systems.
We also proposed possible implementations using photonic crystals,
toroidal microcavities and superconducting systems. Finally we
discussed the capability and advantages of simulating XY spin models
using our scheme and noted the ability of these arrays to simulate
arbitrary quantum networks .

  We acknowledge helpful discussions with A. Carollo and A. Kay.
Also the hospitality of Quantum Information group in NUS Singapore,
and the Kavli Institute for Theoretical Physics where discussions
between DA and SB took place during joint visits. This work was
supported in part by the QIP IRC (GR/S82176/01), the E.U. FP6-FET
Integrated Project SCALA, and an Advanced Research Fellowship from
EPSRC.

\end{document}